\documentclass[aps,prx,twocolumn,
superscriptaddress,
notitlepage,showpacs,floatfix]{revtex4-2}

\usepackage{graphicx}
\usepackage{dcolumn}
\usepackage{bm}
\usepackage{xcolor}
\usepackage{amsmath}
\usepackage{amsfonts}
\usepackage[utf8]{inputenc}
\usepackage{hyperref}
\usepackage{booktabs}
\DeclareUnicodeCharacter{207B}{\textsuperscript{-}}

\begin{document}

\title{QPU-scale randomized benchmarking via Bell-pair injection}

\author{Haripriya Pettugani}
\thanks{These three authors contributed equally to this work.}
\affiliation{Moth, Switzerland}
\affiliation{Center for Quantum Computing and Quantum Coherence (QC2), University of Basel, Switzerland}

\author{María Aguado-Yáñez}
\thanks{These three authors contributed equally to this work.}
\affiliation{Moth, United Kingdom}
\affiliation{ICCMR, University of Plymouth, United Kingdom}

\author{Astryd Park}
\thanks{These three authors contributed equally to this work.}
\affiliation{Moth, United Kingdom}

\author{Daniel Bultrini}
\affiliation{Moth, Switzerland}

\author{James R. Wootton}
\affiliation{Moth, Switzerland}
\date{\today}

\begin{abstract}

Mirror randomized benchmarking (MRB) is an established technique that provides a global error metric at the scale of a whole QPU. To expand upon this we introduce Mirror Quantum Awesomeness (MQA), a hybrid protocol that adds a structured entangling layer to MRB circuits. This enables per-edge correlation dynamics to be tracked via mutual information while preserving the MRB infidelity estimate. The resulting analysis of the injected entangled pairs locates a critical circuit depth, beyond which rudimentary error mitigation techniques can be expected to fail. A topological variant, Topological MQA, supplies a second critical depth via a decoder based on the surface-code decoding problem. Both are validated in simulation and demonstrated on the 156-qubit \texttt{ibm\_fez} and \texttt{ibm\_kingston} processors, where MQA closely agrees with MRB on the entanglement infidelity and the critical depth for \texttt{ibm\_fez} is found to be $\sim 50$.

\end{abstract}

\maketitle

\section{Introduction}\label{sec:intro}

The advent of quantum processing units (QPUs) surpassing 100 qubits marks a pivotal era in computational science, in which quantum systems begin to tackle problems intractable for classical computers, such as simulating molecular interactions or optimizing complex processes \cite{aspuru2005simulated, zhong2020quantum}. The inherent parallelism provided by superposition, combined with interference effects, allows quantum systems to converge efficiently on a solution \cite{nielsen2010quantum}. However, realizing this potential is hampered by inherent errors, crosstalk, coherent noise, drift, and decoherence, which increase with system size and circuit depth, often rendering large-scale computations unreliable \cite{preskill2018quantum}. Traditional metrics, such as gate fidelities, offer localized insights but fail to predict holistic performance, underscoring the need for comprehensive benchmarking that captures how errors manifest in realistic workloads \cite{acuaviva2024benchmarking,hesner2025detector, proctor2025benchmarking}.

Benchmarking serves as a bridge between hardware capabilities and algorithmic utility, providing actionable data for optimization in the noisy intermediate-scale quantum (NISQ) regime \cite{proctor2022measuring, hashim2024practical}. Early protocols focused on customized random circuits to probe entanglement dynamics, but recent innovations emphasize scalability and interpretability. Among these, Mirror Randomized Benchmarking (MRB) \cite{proctor2022scalable} stands out for its ability to estimate average layer infidelities at QPU scales. MRB employs randomized mirror circuit sequences of Clifford layers, combined with Pauli twirls and local Clifford preparations, to measure effective polarization decay and approximate the infidelity of Pauli-dressed \(n\)-qubit layers. Simulations on up to 225 qubits with physically realistic error rates in the range 0.1\%--1\%, together with experiments on up to 16 physical qubits, demonstrate that MRB can detect crosstalk through infidelity growth with qubit count while maintaining robustness under Markovian errors \cite{proctor2022scalable}. However, its reliance on global ensemble averaging leads to high sampling overhead for large systems, limiting fine-grained diagnostics.

Complementing the quantitative focus of MRB is the qualitative accessibility of Quantum Awesomeness (QA) \cite{wootton2018benchmarking, wootton2024quantumpuzzles,moth2025quantumpuzzles}, a benchmarking protocol based on random circuits that generate sets of entangled qubit pairs. QA is designed not only as a quantitative benchmark but also as a visualization-driven tool for qualitative assessment. In this approach, results are presented as puzzles on the device graph, challenging users to infer the underlying structure of qubit pairs as noise progressively increases. The signature metric of QA is the threshold circuit depth at which noise overwhelms the ability of two-body observables to detect correlations, which will be referred to as the `peak of doom'. Experiments on IBM Quantum hardware demonstrate a substantial improvement in this threshold, increasing from approximately four on 16-qubit devices in 2018 \cite{wootton2018benchmarking} to approaching 100 on 156-qubit devices in 2024 \cite{wootton2024quantumpuzzles}.

In both MRB and QA, noise builds up continuously with no notion of a strict critical depth at which the effects of noise dominate the signal and meaningful information cannot be extracted. However, locating such a critical depth on a given device can be valuable, setting a practical ceiling on the circuit depths at which the device can be used productively without massive overheads for error mitigation or correction. The methods presented in this work yield two complementary ways to estimate this critical depth and demonstrate both on IBM Quantum hardware.

Building on these foundations, this work introduces Mirror Quantum Awesomeness (MQA), a hybrid of the above two frameworks. This integrates the creation and analysis of entangled pairs into MRB circuits, retaining the features of MRB while allowing the additional analysis of QA.

This paper is organized as follows. First, the two benchmarking protocols that motivate this work, Mirror Randomized Benchmarking (MRB) and Quantum Awesomeness (QA), are reviewed in Sec. \ref{sec:background}. The proposed Mirror Quantum Awesomeness (MQA) framework is then introduced in Sec. \ref{sec:mqa}, together with its associated benchmarking metrics, which are evaluated through simulation. Next, a topological extension, Topological MQA, is presented in Sec. \ref{sec:topo}, enabling an additional set of correlation metrics. The performance of these metrics is investigated through simulation and compared with those obtained using MQA. Finally, results obtained from experiments on IBM Quantum Processing Units (QPUs) are presented and discussed in Sec. \ref{sec: results}.

\section{Background}\label{sec:background}

\subsection{Mirror Randomized Benchmarking}\label{sec:mrb}

Errors in quantum gates arising from decoherence, imperfect control, and crosstalk accumulate rapidly in multi-qubit circuits, necessitating scalable benchmarking protocols that quantify average performance without exponential resource overheads \cite{krinner2020benchmarking, preskill2018quantum}. Traditional Randomized Benchmarking (RB) methods, such as Standard Randomized Benchmarking (SRB) \cite{magesan2012characterizing, magesan2011scalable} and Direct Randomized Benchmarking (DRB), aggregate diverse error sources into a single infidelity metric by averaging the ensemble over random Clifford circuits. These techniques excel in precision for small systems but falter at scale: SRB requires $\mathcal{O}(n^2 / \log n)$ two-qubit gates per $n$-qubit Clifford to compile \cite{bravyi2021hadamard, aaronson2004improved, patel2003efficientsynthesislinearreversible}, limiting implementations to $n \leq 3$ \cite{mckay2019three}, while DRB, though streamlined, remains confined to $n \leq 5$ \cite{proctor2019direct}. Meanwhile, volume-oriented benchmarks like quantum volume \cite{cross2019validating} and cross-entropy benchmarking \cite{boixo2018characterizing} require classically intractable simulations beyond $n \sim 50$ \cite{boixo2018characterizing, arute2019quantum, cross2019validating}. This scalability chasm obscures holistic insights into many-qubit errors, particularly spatially correlated phenomena like crosstalk, which evade single- and two-qubit calibrations.

MRB was introduced as a protocol that redefines scalable quantum performance assessment, combining conceptual simplicity with broad applicability. MRB employs randomized mirror circuits to enable robust characterization of large-scale quantum processors.

These circuits form a customizable ensemble of Pauli-twirled, self-inverting layers used to estimate the average entanglement infidelity $\epsilon_\Omega = \mathbb{E}_{\Omega}[\epsilon(L)]$ of Pauli-dressed $n$-qubit Clifford layers under a user-specified distribution $\Omega$ over the layer set $\mathcal{L}$. The circuit consists of $d/2$ pairs of layers $L_i$ and $L_i^{-1}$ as shown in the Fig.~\ref{fig:mqa-circuit}, sampled from $\Omega$, interleaved with $d+1$ uniform Pauli layers $P_i$ for twirling and bounded by initial and final single-qubit Clifford layers $F_0$ and $F_0^{-1}$ for local randomization. By construction, ideal circuits yield a single, classically computable bitstring $s_C$, while errors manifest themselves as deviations in the output distributions. Execution involves sampling $K$ circuits per even length $d \geq 0$, running each $N$ times, and computing the effective polarization

\begin{equation}
    S = \frac{4^n}{4^n-1} \left[ \sum_{k=0}^n \left( -\frac{1}{2} \right)^k h_k \right] - \frac{1}{4^n-1},
    \label{eq:effective-polarization}
\end{equation}

where $h_k$ is the probability of Hamming distance $k$ from $s_C$. Fitting $S_d \approx A p^d$ yields $r_\Omega = (4^n-1)(1-p)/4^n \approx \epsilon_\Omega$, correcting for state preparation/measurement (SPAM) errors via binomial inversion. Under Markovian assumptions, exponential decay with $p^2 \approx \mathbb{E}_\Omega[\gamma_{L^{-1}} \gamma_L]$ (where $\gamma = 1 - 4^n \epsilon / (4^n - 1)$) has been theoretically proven~\cite{proctor2022scalable}, with scrambling properties ensuring negligible error correlations. This yields reliable estimates even for non-uniform errors, with relative biases $|\delta_{\mathrm{rel}}| \lesssim 0.2$ in physical regimes.

What elevates MRB is its good scalability and diagnostic power. Simulations on 225-qubit lattices (15$\times$15 grid) with realistic Pauli errors 
(0.1\% single-qubit, 1\% two-qubit infidelities) confirm $r_\Omega \approx \epsilon_\Omega$ across 900 trials, unmasking quadratic crosstalk scaling absent in gate-local models. On IBM Quantum devices (Quito and Rueschlikon), MRB benchmarks $n \leq 16$ qubits, matching DRB for $n \leq 5$ while exposing $n$-dependent divergences from calibrations quantifying crosstalk contributions invisible to isolated RB \cite{proctor2022scalable}.

\subsection{Quantum Awesomeness}\label{sec:qa}

Quantum Awesomeness (QA) \cite{wootton2018benchmarking,wootton2024quantumpuzzles} is a complementary benchmarking framework based on random circuits, proposed initially to visualize and interpret the behavior of quantum devices through human-interpretable correlation patterns. In its original formulation, QA applies random entangling layers followed by reconstruction attempts based on the observed output statistics. The protocol evaluates how effectively these correlations persist as the circuit depth increases, thereby providing an intuitive measure of the buildup and noise accumulation.

In QA, the intended output state of the computation is always a set of entangled pairs of the form $\cos\theta_{jk} |00\rangle + \sin\theta_{jk} |11\rangle$, where the value of $\theta_{jk}$ varies from pair to pair. As such, $\langle Z_j \rangle = \langle Z_k \rangle$ for the two qubits of each pair, and $\langle Z_j Z_k \rangle =1$. These theoretical values then provided the baseline for analyzing results in practice. Analyzing the degradation of these correlations with increasing circuit depth allows the transition from coherent quantum dynamics to noise-dominated behavior to be determined. This focus on the structure of correlations between qubit pairs is the major difference between QA and MRB.

A notable feature of the QA framework is its gamified interpretation~\cite{wootton2018benchmarking,moth2025quantumpuzzles}. Measurement statistics from each qubit are mapped to numbers or colors, and the challenge is to identify which qubits formed entangled pairs in the most recent circuit layer. Essentially, players are presented the $\langle Z_j \rangle$ values and, using the fact that $\langle Z_j \rangle = \langle Z_k \rangle$, are asked to identify the pairs. At shallow circuit depths, this puzzle is straightforward, since correlations are strong and localized. As the circuit depth increases, the spread of entanglement, crosstalk, and decoherence gradually obscures the correlation structure, making the puzzle progressively more difficult. The point at which meaningful structure disappears corresponds operationally to the onset of chaotic circuit behavior and the breakdown of reversibility. This frame of reference provides an intuitive visualization of the loss of quantum coherence and the buildup of noise.

\subsection{Motivation for a Hybrid MQA-MRB Framework}

MRB and QA emphasize different aspects of the behavior of quantum processors. MRB provides scalable, quantitatively rigorous fidelity estimates based on circuit reversibility, whereas QA offers a qualitative, human-interpretable view of the correlation structure and its degradation. These complementary strengths motivate a hybrid framework that combines the structured reversibility of MRB with the additional quantitative and qualitative insights of QA.

In this work, Mutual Information (MI) is employed as the primary metric to quantify qubit correlations. Compared with the original probability-based pairing metric used in Quantum Awesomeness (QA), MI offers a more comprehensive characterization of quantum correlations by enabling robust identification of noise patterns, particularly in deeper and more complex circuits.

Building on this advantage, the proposed Mirror Quantum Awesomeness (MQA) protocol thus integrates the strengths of Mirror Randomized Benchmarking (MRB) with the enhanced QA methodology. The resulting hybrid circuit architecture combines QA-inspired entangling layers, MRB mirror blocks, and mutual-information-based analysis to investigate the evolution of correlation dynamics throughout circuit execution.

A central motivation of this work is to build on the `peak of doom' present in QA, which informally can be used to identify the circuit depth at which noise begins to dominate the structured correlations the protocol relies on. We show that MQA can be used to provide a more concrete manifestation of such as peak (Sec.~\ref{sec:peak}), and that Topological MQA provides an additional perspective inspired by QEC thresholds (Sec.~\ref{sec:topo}). Both estimators are studied across a range of depolarizing-noise strengths in simulation and then applied to IBM Quantum hardware.

\section{Mirror Quantum Awesomeness}\label{sec:mqa}

\subsection{Structure of the Circuit}\label{sec:structure}

In the MRB framework, circuits are constructed from alternating layers of random single-qubit Pauli gates and randomly sampled Clifford-type circuit layers. The Pauli layers act as a twirling mechanism, introducing strong local randomization that prevents coherent errors from accumulating constructively over repeated circuit depth. As a result, coherent error contributions are effectively converted into stochastic-like behavior, leading to more stable decay curves and simplifying the interpretation of benchmarking data.

A width-$w$ randomized mirror circuit of benchmark length $d$ typically consists of the following components:
\begin{enumerate}
    \item An initialization layer that prepares all qubits in the computational basis state $|0\rangle^{\otimes w}$.
    \item A sequence of layers sampled from a specified distribution $\Omega$, where each repeated unit consists of:
    \begin{enumerate}
        \item a uniformly random single-qubit Pauli layer, followed by
        \item a circuit layer sampled from $\Omega$, which may contain both single-qubit Clifford gates and multi-qubit entangling operations.
    \end{enumerate}
    \item A central random Pauli layer applied before inversion.
    \item A mirrored inverse sequence, where the sampled circuit layers are applied in reverse order and replaced by their inverses, while the Pauli layers are resampled independently.
    \item A final closing Pauli layer.
    \item Measurement in the computational basis.
\end{enumerate}
The length of the circuit corresponds to the depth of two-qubit gates. Given this condition, the terms `length' and `depth' can be used interchangeably.

Note that the original MRB protocol~\cite{proctor2022scalable} additionally wraps the Pauli-and-sampled-layer sequence in an initial layer $F_0$ of uniformly random single-qubit Clifford gates and its inverse $F_0^{-1}$ at the end. Our implementation builds on \texttt{qiskit-device-benchmarking}~\cite{qiskit-device-benchmarking}, which permits these layers to be omitted, and we take that option throughout. The rationale and its effect on fit quality are discussed in Section~\ref{sec: results}.

\begin{figure*}[t]
    \centering
    \includegraphics[width=0.9\textwidth]{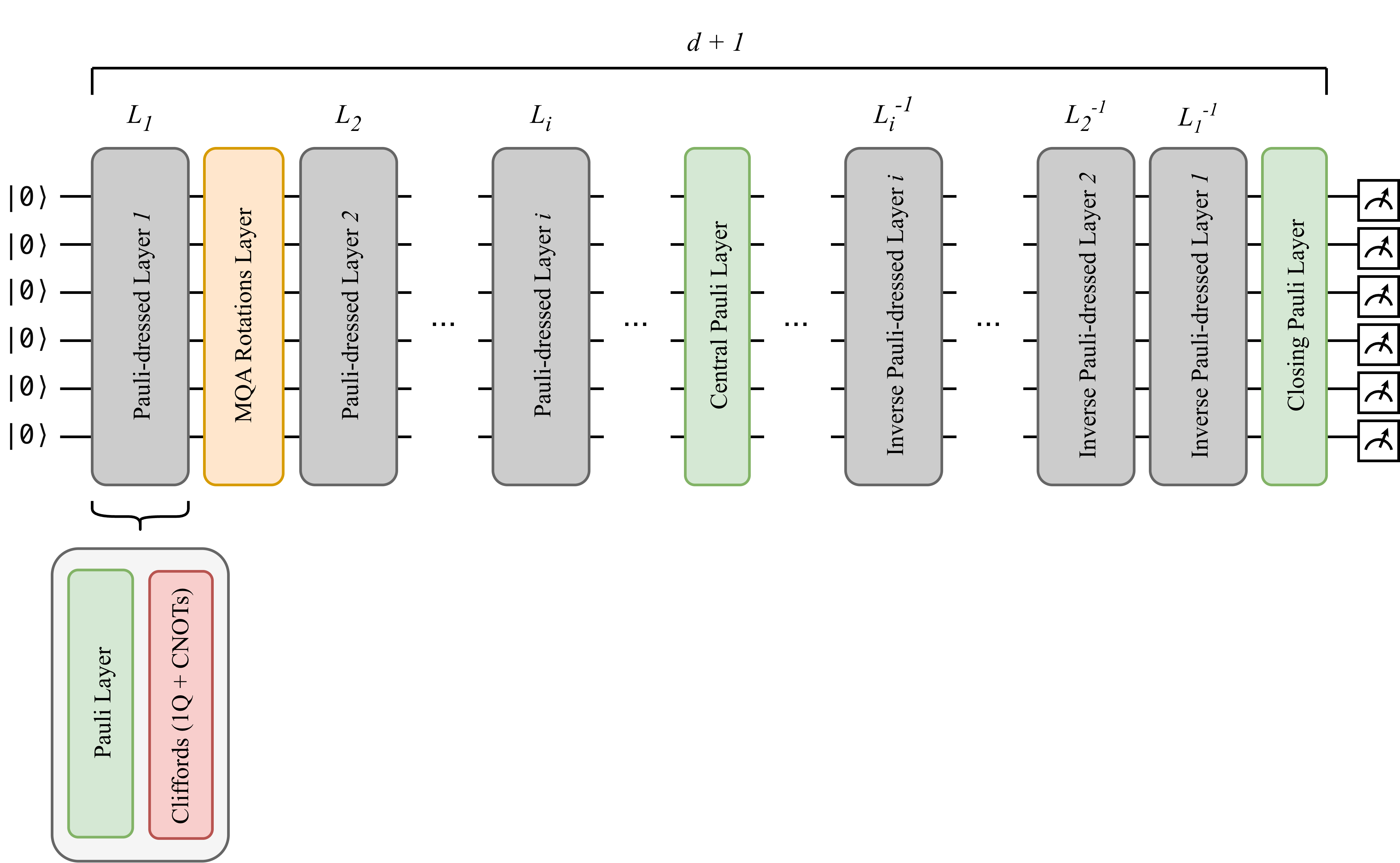}
    \caption{\textbf{MQA Circuit Structure.} The yellow block highlights the additional structured layer ($\mathrm{R_x}(\theta)$ on control qubits of the first entangling layer) inserted in the forward half. Except for the initial and final single-qubit Clifford layers (omitted), all other layers follow the standard MRB architecture.}
    \label{fig:mqa-circuit}
\end{figure*}

This construction ensures that the forward randomized circuit evolution is followed by a carefully constructed mirror portion, such that in the absence of noise the overall circuit ideally returns the system to its initial state, up to a string of Paulis.

Although MRB provides a scalable benchmarking backbone, it does not explicitly enforce tunable correlations or controlled entanglement structure across different circuit depths. To address this limitation, an additional structured rotation layer is incorporated into the MRB architecture, resulting in the modified circuit family referred to as MQA.

As illustrated in  Fig.~\ref{fig:mqa-circuit}, the MQA circuit preserves most of the MRB mirror structure, consisting of a sequence of Pauli-dressed circuit layers, their corresponding inverse layers, and a final closing Pauli layer. The key modification is the insertion of an intermediate rotation layer within the forward half of the circuit, specifically between the first and second Pauli-dressed layers.

The inserted MQA rotation layer consists of single-qubit $R_x(\theta)$ rotations applied on the control qubits of all qubit pairs involved in the first layer controlled-NOT (CX) operations. The $CX$ gates of the first layer and its mirror then conjugate these rotations, yielding an entangling $R_{xx}(\theta) = \exp(-i \frac{\theta}{2} X \otimes X)$ operation. This therefore has the effect of injecting entangled pairs into the circuit.

By varying the rotation angle $\theta$, the degree of correlation introduced into the circuit can be systematically controlled. In particular, choosing $\theta = \pi/2$ generates maximally entangled pairs, establishing strong correlations at the outset, whereas $\theta = 0$ introduces no additional entanglement and reduces the circuit back to the original MRB structure.

Intermediate values of $\theta$, such as $\theta = \pi/4$, generate partially entangled states and enable exploration of regimes beyond purely Clifford dynamics. This is significant since it extends the benchmarking family to include controlled non-Clifford perturbations, allowing a more comprehensive characterization of error propagation and correlated noise effects under general quantum evolution.

\subsection{Circuit Parameters and Modified Design}\label{sec:parameters}

The MQA framework extends the mirror-circuit benchmarking protocol by refining several core parameters that control the circuit structure, data sampling, and analysis. The following subsections describe the primary parameters and methodological differences introduced relative to MRB.

\subsubsection{Entangling Angle}

The entangling angle defines the rotation applied in two-qubit gates. For Clifford circuits, primarily $0$ and $\pi/2$ are used. The zero rotation generates no entanglement, while $\pi/2$ produces a maximally entangled pair. In MQA, these angles were applied explicitly in the first layers of the circuits to control entanglement and facilitate simulations. The protocol specifically inserts $RX$ gates into the control qubits of any qubit involved in a $CX$ in the first layer, leading to entanglement between all pairs in that layer.

\subsubsection{Two-Qubit Gate Density} 
A key feature of MRB circuits is the density of two-qubit entangling gates, since these gates typically dominate noise in near-term quantum devices. For a circuit $C$ of width $w$ and benchmark depth $d$, let $\alpha$ denote the total number of two-qubit gates. The two-qubit gate density is defined as

\begin{equation}
\xi = \frac{2\alpha}{wd}.
\end{equation}

This can be interpreted by viewing the circuit as a $w \times d$ grid, where each two-qubit gate occupies two qubit-locations. For example, a density value of $0.25$ implies that approximately one-quarter of all possible pairs of two-qubits are selected to form entanglement gates in each layer, which in practice results in about half of the qubits being involved in entanglement at a given time, providing a balanced trade-off between circuit depth and entanglement complexity.

In this work, the default density 0.25, as described in the MRB \cite{proctor2022scalable}, is used in most of the analysis, but is later modified to study the effect of circuit topology. Controlling $\xi$ allows one to tune circuit complexity and entanglement while maintaining high disorder, which is crucial for benchmarking quantum devices.

\subsubsection{Sampling Algorithm}

Randomized mirror circuits are generated by sampling layers from a defined set $\mathcal{L}$ of $n$-qubit operations.  
Here, $\mathcal{L}$ consists of all parallel applications of CNOTs between connected qubits and all 24 single-qubit Clifford gates.  
The choice of sampling algorithm determines the structure and density of two-qubit gates within each layer, directly influencing circuit depth, entanglement structure, and error sensitivity.  
In this work, two types of layer samplers were employed: (i) the \textit{Edge-Grab Sampler}, and (ii) the \textit{Matching Sampler}.

\begin{enumerate}
    \item \textbf{Edge-Grab Sampler.} 
The placement of two-qubit gates in randomized mirror circuits is determined using an \emph{edge-grab sampler} parameterized by a target expected two-qubit gate density $\bar{\xi}$. The sampler selects a set of disjoint edges from the device connectivity graph, ensuring that no qubit participates in more than one two-qubit gate within the same layer.

The sampling procedure is defined as follows:

\begin{enumerate}
    \item Initialize a candidate edge set $E=\emptyset$, and let $E_r$ be the set of all edges in the hardware connectivity graph.
    \item While $E_r$ is not empty:
    \begin{enumerate}
        \item Select an edge $v$ uniformly at random from $E_r$.
        \item Add $v$ to $E$, and remove from $E_r$ all edges that share a qubit with $v$.
    \end{enumerate}
    \item Each edge in $E$ is then included in the final sampled edge set independently with probability
    \begin{equation}
        p = \frac{w\bar{\xi}}{|E|},
    \end{equation}
    where $w$ is the circuit width and $|E|$ is the size of the candidate set.
\end{enumerate}

This method produces layers with a controllable expected number of two-qubit gates, allowing the overall circuit ensemble to achieve an average two-qubit gate density close to $\bar{\xi}$. The edge-grab sampler therefore provides a scalable and hardware-compatible approach for generating randomized circuit layers with tunable entangling complexity.

    \item \textbf{Matching Sampler.} 
    The matching-based sampling algorithm is constructed using a maximum-weight matching approach and was primarily used for analyzing device topology.  
    Given a set of $w$ qubits, their connectivity graph, and a desired two-qubit gate density $\xi_s$, this sampler proceeds as follows:
    \begin{enumerate}
        \item Perform a maximum-weight matching $E$ on the connectivity graph, assigning random weights to edges to introduce stochasticity while ensuring that $E$ consists of disjoint connected qubit pairs.
        \item From the matched edge set $E$, select each edge with probability
        \begin{equation}
            p = \frac{w\xi_s}{2|E|},
            \label{eq:edge-probability}
        \end{equation}
        
        where $|E|$ is the number of matched edges.  
        The selected edges represent the two-qubit operations for the given layer.
        \item Assign random single-qubit Clifford gates to all remaining qubits not participating in two-qubit operations.
    \end{enumerate}
    This generates layers with an expected two-qubit gate density of $\xi = \xi_s/2$.  
    When these layers are interleaved with single-qubit Pauli layers, the overall expected density converges to $\xi_s$ for large circuit sizes.  
    In this work, the matching sampler was specifically used for topology analysis, focusing on loop and square configurations.
\end{enumerate}

\subsubsection{Parity-Based Measurement Analysis}

As discussed in Sec. \ref{sec:mrb}, the effective polarization, $S$, used as the basis of the analysis of MRB uses the Hamming distance from the expected output bit string $s_C$. For MQA there is no such bit string, since the expected outcome is instead a set of Bell pairs. The exact form of the Bell pairs will also depend on the interleaved Paulis in the circuit. Though this expected outcome can still be efficiently computed, there is nevertheless a need to determine how to calculate $S$.

One option is to use the value of $s_C$ that would arise if the $RX$ gates were not present. With this strategy, these gates are effectively treated as an initialization error when calculating $S$. Since randomized benchmarking is able to extract information about errors in a gate sequence independent from SPAM errors, this should still allow the correct results to be determined. However, the large amount of effective additional noise will increase the statistical overhead required for good results.

Instead, a truncated string $s'_C$ is calculated. For all qubits not involved in pairs, the corresponding bit value of $s_C$ is present unchanged in $s'_C$. For each pair of qubits that are entangled due to the addition of the $RX$ gates, the two bits of $s_C$ are replaced by their parity in $s'_C$. All outcome bit strings are similarly truncated, with pairs of bits corresponding to entangled pairs being replaced by their parity. The analysis of Eq. (\ref{eq:effective-polarization}) to calculate $S$ can then proceed as normal, with the Hamming distance defined on the truncated strings.

\subsection{Mutual Information}\label{sec:mi}

In QA, each pair is prepared with a different entangling angle~\cite{wootton2018benchmarking}. This means that $\langle Z_j \rangle = \langle Z_k \rangle$ when the two qubits are part of the same pair, but $\langle Z_j \rangle \neq \langle Z_k \rangle$ otherwise. These properties were then used to define the analysis of the effects of noise in QA.

In MQA however, we will consider the case that all entangled pairs share the same angle. The metrics of QA will therefore not be directly applicable, since all entangled qubits will share the same $\langle Z_j \rangle$. To better identify the pairs we must therefore consider two qubit metrics.

Specifically, we use the Mutual Information (MI) calculated via the Shannon entropies, denoted with $I$, which quantifies the shared information between two random variables~\cite{nielsen2010quantum}. In our case the random variables are the probability of an outcome \texttt{0} or \texttt{1} for two qubits $j$ and $k$. For a given pair of qubits, where the marginal probability distributions for measurement outcomes are $p_j$ and $p_k$, and the joint distribution is $p_{j,k}$, $I$ is defined in terms of the Shannon entropy $H(p)$ as

\begin{equation}
    I(j;k) = H(p_j) + H(p_k) - H(p_{j,k}).
\end{equation}

\begin{figure}
    \centering
    \includegraphics[width=0.7\linewidth]{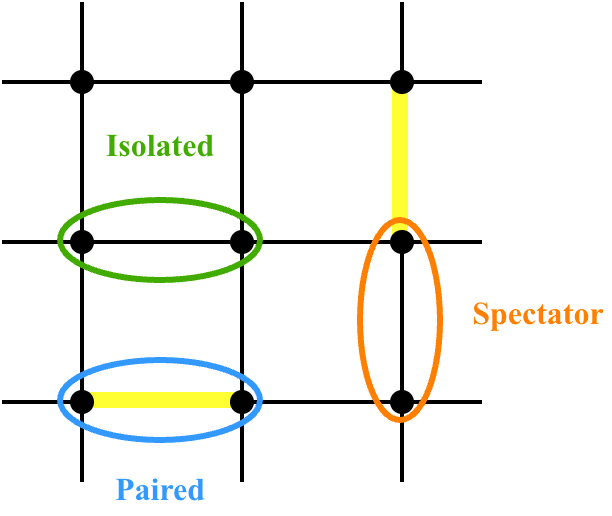}
    \caption{\textbf{Types of Edges}. The yellow lines indicate qubit pairs that become entangled by the newly inserted MQA layer. Blue edges represent paired qubits, orange edges are the spectator qubits, and the remaining isolated edges are shown in green.}
    \label{fig:edge-types}
\end{figure}

To better understand how these correlations arise and evolve, qubit pairs are classified into three distinct categories, as shown in Fig.~\ref{fig:edge-types}:

\begin{enumerate}
    \item \textbf{Paired Edge}: Qubit pairs selected to be entangled by the newly inserted MQA layer.
    \item \textbf{Spectator Edge}: Qubit pairs in which at least one qubit is connected to a paired qubit but is not itself part of an entangling operation.
    \item \textbf{Isolated Edge}: Qubit pairs whose qubits are neither involved in nor adjacent to any paired edge.
\end{enumerate}

To characterize the global behavior of each edge category, the MI is averaged over all qubit pairs within that category. We refer to this as the Mean Mutual Information (MMI): the average over pairs to yield a holistic value for each of the three categories. When $\theta=\pi/2$ and edges are maximally entangled, the MMI takes the value $1$ for paired edges, while the spectator and isolated edges will be $0$. These will be the approximate values at the beginning of the MQA process, when the depth is low.

\subsection{Peak of Doom}

As depth increases, noise begins to disrupt the correlations of paired edges. This can lead to degradation of correlations as they become obscured by noise. It can also lead to the generation of novel correlations due to the mirrored structure. Specifically, a random $X$ error on the control qubit of a $CX$ in any entangling layer is conjugated by both that $CX$ and that from its corresponding inverse layer. This results in it becoming a random $X \otimes X$ error. Noise-induced correlations can therefore arise for the pairs of qubits from the second and later layers. This can generate correlations in spectator and isolated edges. Such correlations will also occur in the first layer if $\theta =0$, or is otherwise small enough that noise-induced correlations arise faster than entanglement decays.

Due to these effects, the increase in the depth and corresponding increase in noise will mean that the MMI for the entangled edges will decay, while the MMI for the spectator and isolated edges will initially grow. However, since increasing noise is always destructive to correlations, the MMI for these edges will reach a peak and then drop again.

This peak marks a notable point in the circuit. Prior to the peak, the structure of the circuit is dominant, with this structure modifying noise to create noise-induced correlations. After the peak, it is instead the effects of noise that are dominant in obscuring the structure of the circuit. The peak can therefore serve as a measure of the depth at which the regime for which extraction of a clear signal is straightforward, to that for which much more in-depth error mitigation would be required.

A qualitatively similar peak was also present in QA. Given the outreach-focused nature of that work, it was referred to as the `peak of doom'. For consistency, we will use the same term for the peak of MQA.

\begin{figure}[!htbp]
    \centering
    \includegraphics[width=1\linewidth]{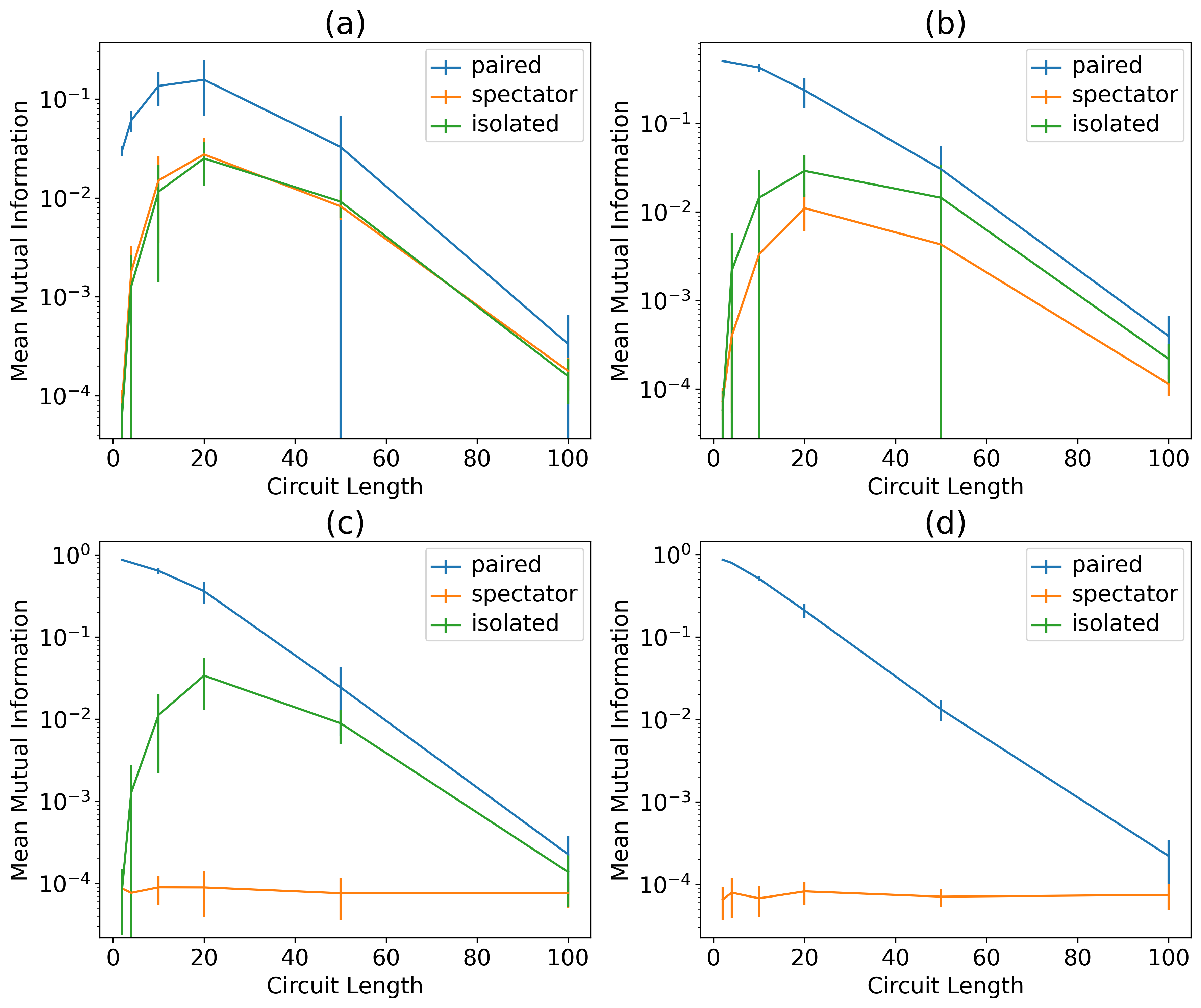}
    \caption{\textbf{MMI for Custom Noisy Backend}. Dependence of the Mean Mutual Information (MMI) on the initial entangling angle $\theta$ and entanglement density $\xi$ for each edge type in an 8-qubit system, with $p_2=0.01$, \texttt{samples = 20} and \texttt{shots = 10000}. (a) $\theta=0$, $\xi=0.25$. (b) $\theta=\pi/4$, $\xi=0.25$. (c) $\theta=\pi/2$, $\xi=0.25$. (d) $\theta=\pi/2$, $\xi=1$.}
    \label{fig:angles-densities}
\end{figure}

Fig.~\ref{fig:angles-densities} shows the behavior of the different edge types for varying rotation angles and entangling densities, evaluated on a custom noisy backend simulator in order to better understand the observed decay. This backend is implemented as a manually defined subclass of Qiskit’s \texttt{GenericBackendV2}, allowing control over the number of qubits, the set of basis gates, the connectivity, and the error rates assigned to one-qubit and two-qubit operations. In this model, the one-qubit error rate, $p_1$, is defined relative to the two-qubit error rate, $p_2$, as $p_1 = p_2/10$. By overriding the default parameters of \texttt{GenericBackendV2}, this custom backend provides a controlled environment for benchmarking MQA routines.

To study angles other than multiples of $\pi/2$, an efficient stabilizer backend cannot be used. A system of eight qubits is therefore used for simplicity. For the case of $\theta = \pi/2$, the paired edges become maximally correlated. This leaves no room for correlations to build on neighboring spectator edges, so the MMI of the spectator edges remains pinned at zero. For $\theta = \pi/4$, there is room for noise to build correlation on the spectator edges, which is why their MMI shows a peak followed by a decay. In contrast, the isolated edges always exhibit a peak regardless of the initial entangling angle, since they are never confined to entangled states, and noise can always build correlations on them. When $\theta = 0$, both the spectator and paired edges effectively behave like isolated edges, showing a peak and returning to the MRB-like behavior where no additional rotation gates are applied. Although all three curves show a peak, only the spectator and isolated curves overlap, while the paired edges remain at higher values. This is due to the correlations induced by the CNOTs, which are only applied on the paired edges. The two-qubit gate density $\xi$ also influences how correlations develop. At maximal density, no isolated edges exist because the distribution of entangling gates is saturated. Reducing the density below this maximum reintroduces the isolated-edge curve.

\begin{figure}[!htbp]
    \centering
    \includegraphics[width= \linewidth]{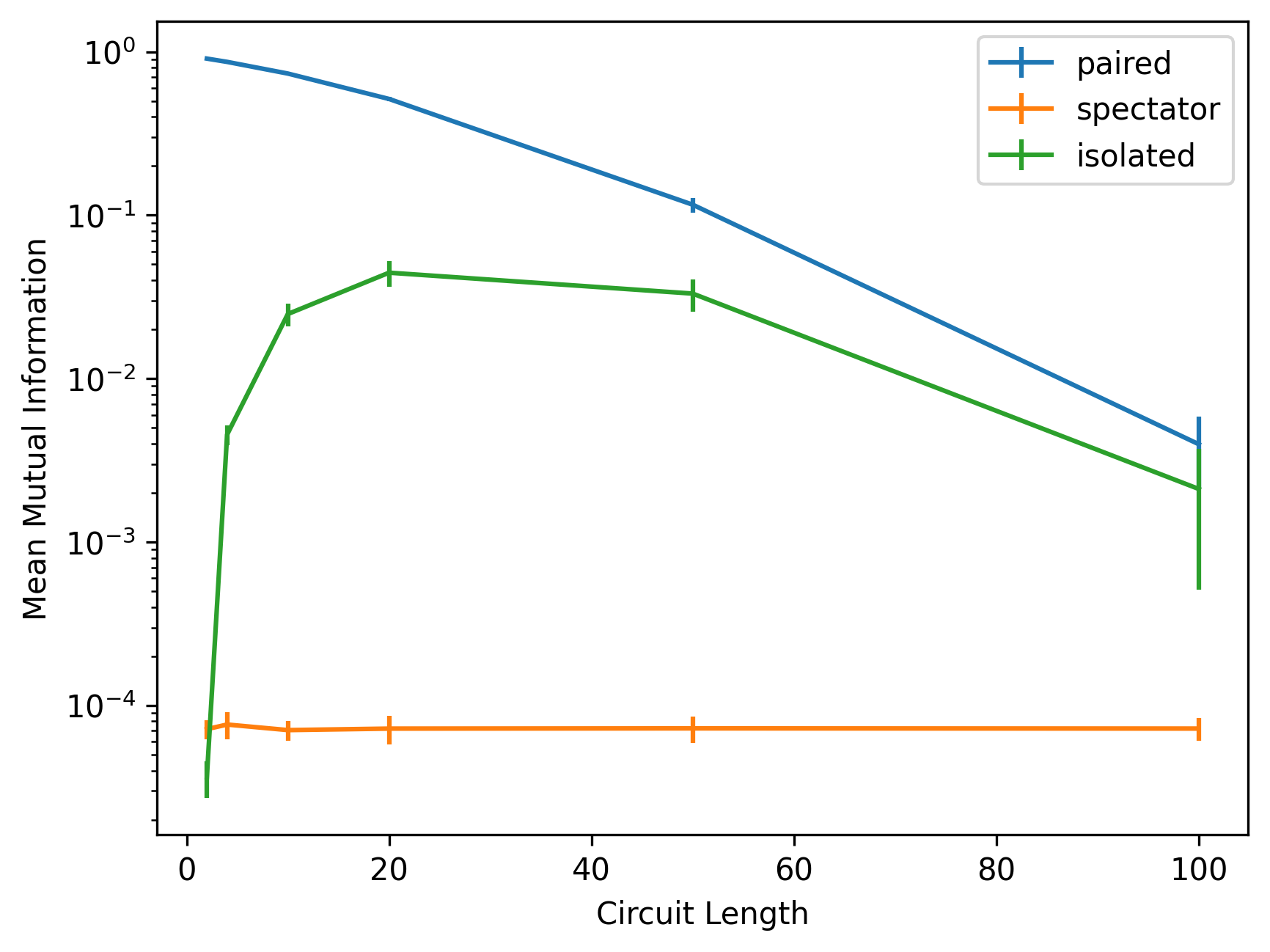}
    \caption{\textbf{MMI for Stabilizer Backend}. MQA simulations of 156 qubits on the \texttt{ibm\_fez} topology with $p_2=0.01$, \texttt{samples = 20} and \texttt{shots = 10000}. The initial entangling angle is fixed at $\theta = \pi/2$, consistent with the stabilizer backend that allows only Clifford gates.}
\label{fig:mqa_stab_fez_whole_pi_over_2}
\end{figure}

Fig.~\ref{fig:mqa_stab_fez_whole_pi_over_2}
shows results from a 156 qubit simulation of the full \texttt{ibm\_fez} coupling map. This requires a stabilizer simulator, meaning that the entangling angle must be set to $\pi/2$. The strong correlations on the paired edges leave no room for correlation build-up on the spectator edges, mirroring the behavior observed in the custom noisy backend. As expected, the isolated edges still exhibit a peak.

\begin{figure}[!htbp]
    \centering
    \includegraphics[width= \linewidth]{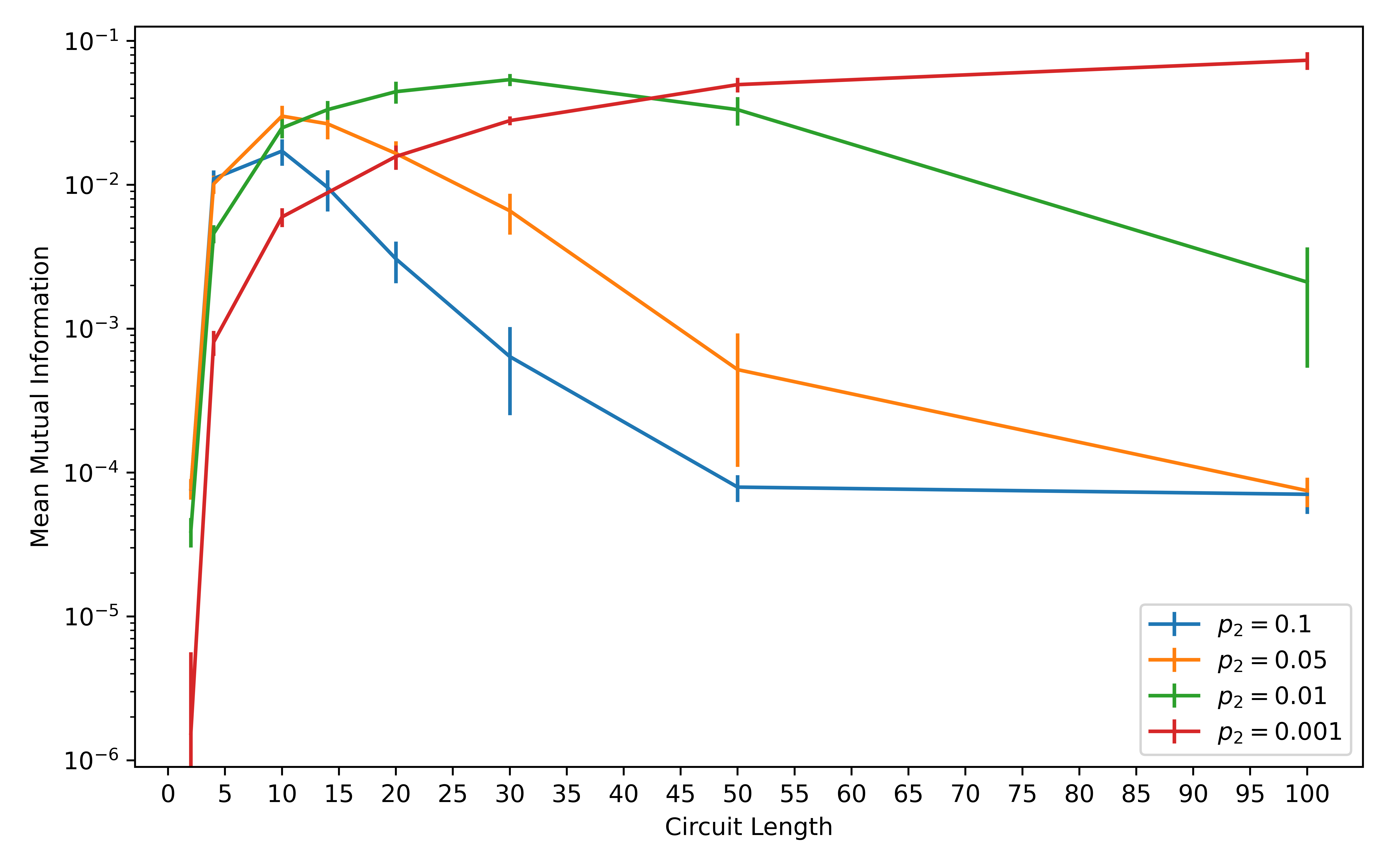}
    \caption{\textbf{Isolated Pairs for Stabilizer Backend}. MQA simulations of 156 qubits on the \texttt{ibm\_fez} topology for isolated pairs, showing the impact of depolarizing noise on the MMI evolution. The initial entangling angle is fixed at $\theta = \pi/2$, with \texttt{lengths} = [2,4,10,14,20,30,50,100], \texttt{samples = 20} and \texttt{shots = 10000}.}
\label{fig:peak_shift_comparison}
\end{figure}

The depth at which the MMI peaks depends on the depolarizing noise strength, as shown in Fig.~\ref{fig:peak_shift_comparison}. Here only the MMI for isolated edges is shown for the 156 qubits of \texttt{ibm\_fez}. As the error rate $p_2$ increases, the peak shifts to smaller depths, since stronger noise accelerates both the growth and the decay of the MMI curve.

Specifically, we see that a $0.1\%$ error rate results in a peak that is not yet resolved even at a depth of $100$. For $p_2=1\%$ the peak is around depth $30$. For both $p_2=5\%$ and $p_2=10\%$ the peak is at around depth $10$. These values can then be used to benchmark results from real QPUs against this standardized noise model.

\section{Topological MQA}\label{sec:topo}

In the methods considered so far, there is no requirement that every qubit is paired in each entangling layer. This makes the methods highly flexible, allowing them to adapt to every coupling graph. However, it also prevents the possibility of analyzing device-wide topological properties. As such, we will now consider a restricted variant, \emph{Topological MQA} (TMQA), which supports a novel metric inspired by decoding of topological quantum error correcting codes.

TMQA assumes that the coupling graph $G$ (i) admits a perfect matching, so that every qubit can be paired with exactly one neighbor, and (ii) is embeddable in two dimensions with two distinguished disjoint boundaries that can be labeled \emph{left} and \emph{right}. These are chosen such that they lie at maximum graph distance from one another. For the square-lattice coupling maps used in practice, these boundaries are the qubit columns at the two opposite ends of the lattice.

Given such a graph, two puzzle variants are then defined: a \emph{closed} variant in which every qubit is paired, and an \emph{open} variant in which a single qubit on each boundary is left unpaired while all other qubits are paired. Any pairing of the former type then necessarily differs from one of the latter type by at least a path from left to right along which the pairings are complementary. This enforces a minimum difference between such pairings that increases with the code width, $L$. 

A decoder, or \emph{bot}, plays the game by inferring the most likely pairing from the measured mutual-information matrix. The bot solves a minimum-weight perfect matching problem on the coupling graph, with edge weights derived from the per-edge MI values so that pairs with high MI receive low cost. This matching process is modified to allow for either a fully-paired solution (closed), or one with a single isolated qubit on either side (open). The bot's final output is then reduced to a single binary classification of which of these topologically distinct forms of solution is optimal.

More concretely, the matching for $P_{\mathrm{pairs}}$ runs on the QPU's coupling graph directly. The matching for $P_{\mathrm{topo}}$ runs on the same graph augmented with two \textit{fake} qubits: one connected to every qubit on the left boundary, the other to every qubit on the right, together with an edge between the two fakes; all of these added edges carry zero weight. Solutions in which the two fake qubits pair with each other correspond to the closed variant, since every genuine qubit must then be paired with another genuine qubit. Solutions in which each fake qubit pairs with a genuine qubit correspond to the open variant, with one boundary qubit on each side excluded from the genuine pairing.

This single bit of information, whether the pairing is closed or open, is topologically protected by the width of the graph: To convert the closed solution into the open one, the bot's matching must be flipped along an entire connecting path, so the closed-vs.-open label is encoded redundantly along that path and cannot be flipped by any local noise process. This parallels the encoding of a logical bit in a topological quantum error-correcting code such as the surface code~\cite{Dennis2002}, where the logical operator is similarly a path connecting opposing boundaries. By analogy with QEC threshold behavior, one therefore expects the probability of correct decoding to remain close to unity at shallow circuit depth and to drop sharply toward chance once the per-pair correlations become difficult to deduce, with the transition sharpening as the distance $L$ grows.

To probe this behavior, two quantities are tracked as functions of circuit depth: $P_{\mathrm{topo}}$, the probability that the bot correctly classifies the puzzle as closed or open, and $P_{\mathrm{pairs}}$, the probability that the bot identifies all pairs correctly.

\begin{figure}[!htbp]
    \centering
    \includegraphics[width=0.9 \linewidth]{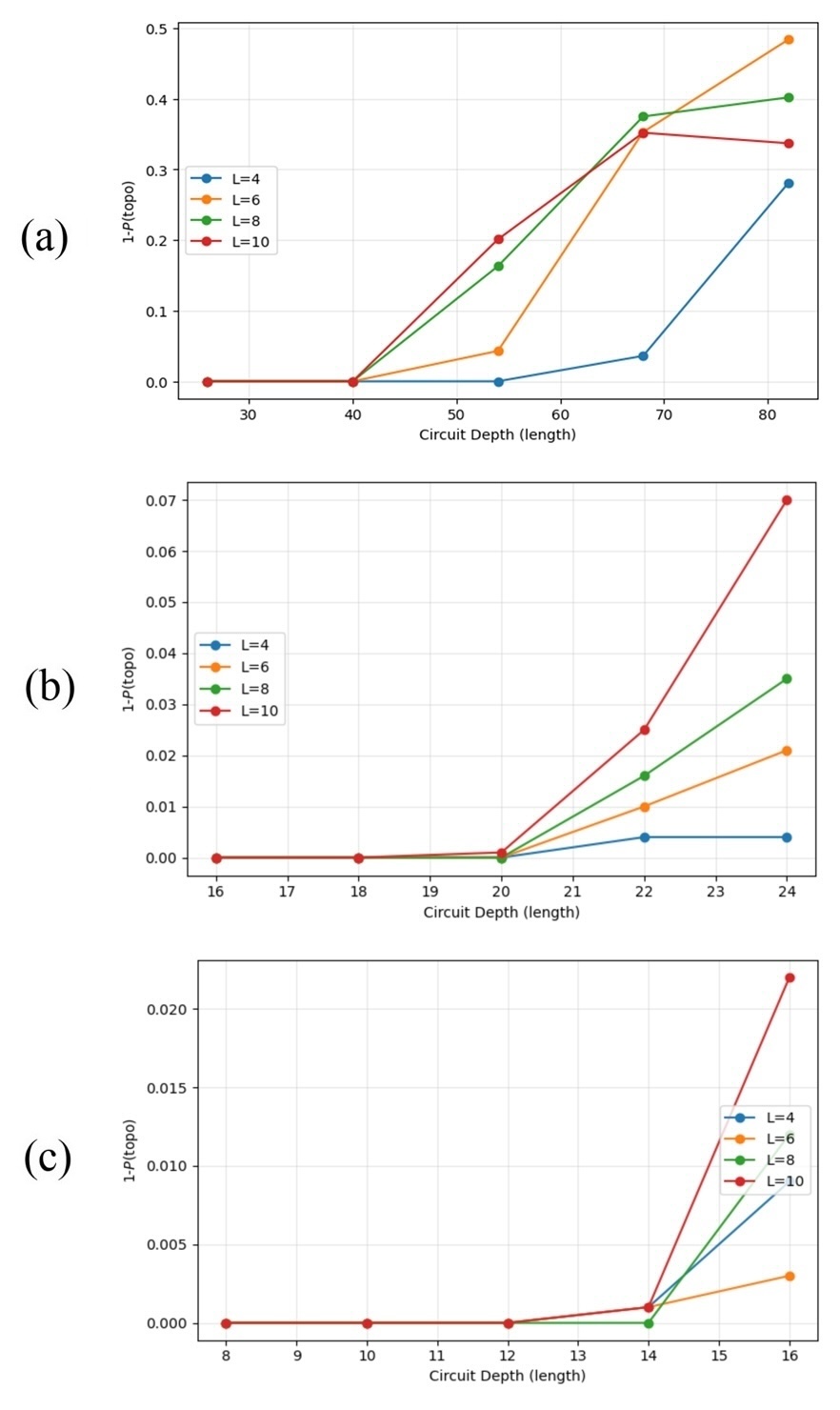}
    \caption{\textbf{$P_{topo}$ for Custom Noisy Backend}. Simulations of TMQA on square lattices of different sizes. (a) $p_2=0.01$. (b) $p_2=0.05$. (c) $p_2=0.1$.}
\label{fig:tmqa_sim_topo}
\end{figure}

\begin{figure}[!htbp]
    \centering
    \includegraphics[width=0.9 \linewidth]{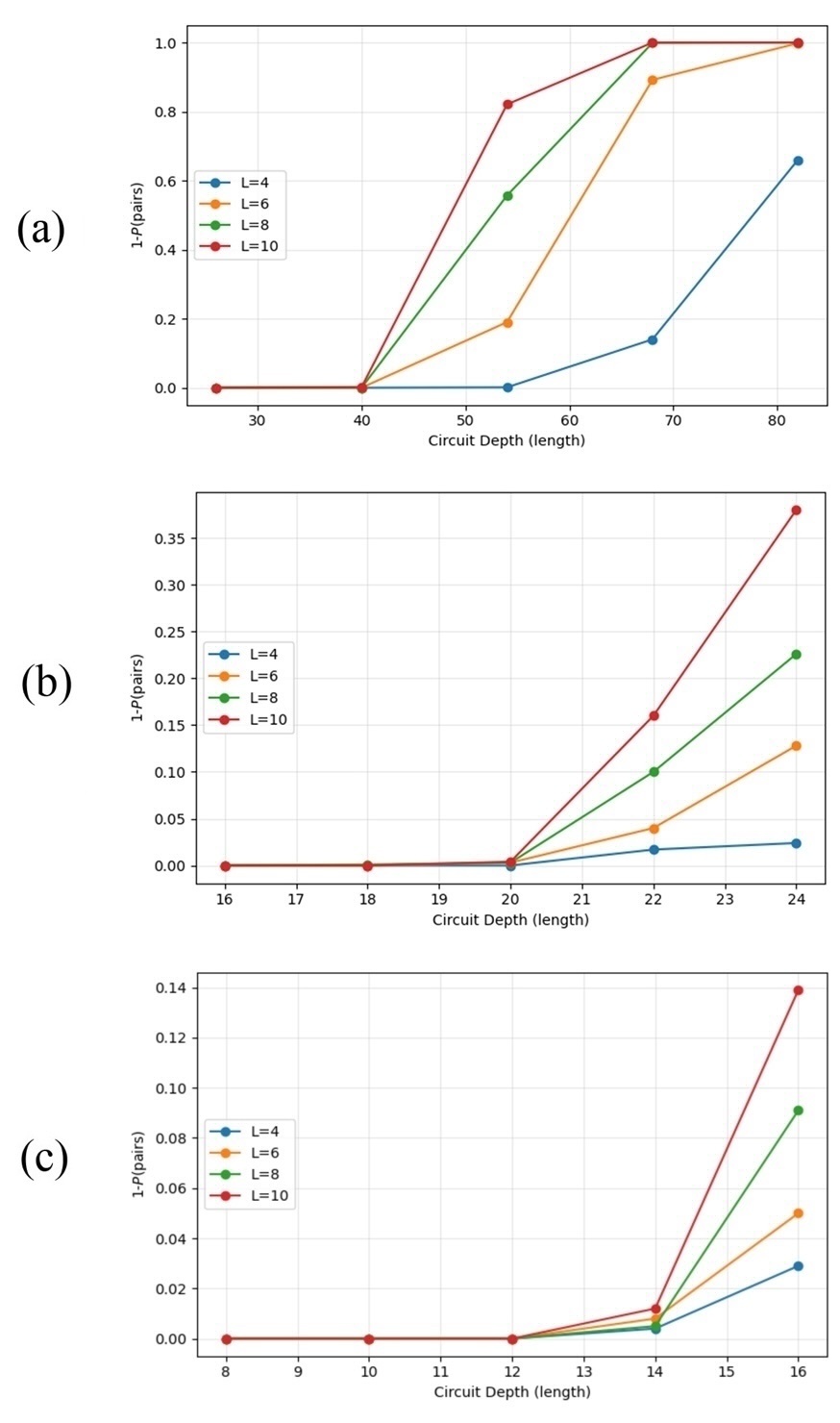}
    \caption{\textbf{$P_{pairs}$ for Custom Noisy Backend}. Simulations of TMQA on square lattices of different sizes. (a) $p_2=0.01$. (b) $p_2=0.05$. (c) $p_2=0.1$.}
\label{fig:tmqa_sim_pairs}
\end{figure}

Results from simulations for a range of grid sizes are shown in Fig.~\ref{fig:tmqa_sim_topo} for $P_{topo}$ and Fig.~\ref{fig:tmqa_sim_pairs} for $P_{pairs}$. Distinct threshold behavior can be seen in both cases. Note that since each run of $P_{topo}$ yields only a single bit (success or failure for the bot) statistical noise results in less distinct separation of different system sizes. Prior to a critical depth, the probability of misidentifying the topological class is practically zero. Afterwards the probability climbs sharply, with increasing gradient for increasing $L$. Note that, in contrast to similar behavior for QEC thresholds, the crossing point occurs close to the x axis rather than in the middle of the plot.

Again, these results allow us to easily see the critical circuit depth at which the effect of noise becomes overwhelming. Specifically, it lies at a depth of around $12$ for $p_2=10\%$, $20$ for $p_2=5\%$ and $40$ for $p_2=1\%$. The threshold depth for $p_2=0.1\%$ could not be resolved even up to a depth of 120.

These results, though obviously not identical, are broadly similar to those identified by the peak of doom with MQA. They therefore represent different means by which we can determine a guideline depth up to which noise can be mitigated, and beyond which mitigation can be expected to be difficult. Indeed, the TMQA case can be considered to be a form of QEC-inspired mitigation, with which a circuit encodes a single bit of information in a topologically protected manner, and mitigation via post-processing is used to retrieve the bit. The critical depth is exactly that for which this method becomes impossible. We therefore propose that these methods can provide a guide to the circuit depth at which error mitigation methods without an increased circuit execution overhead will fail. However, mitigation methods with such an overhead, such as probabilistic error cancellation~\cite{vandenberg2023pec}, may still provide good results at higher depth.

\section{Results}\label{sec: results}

This section presents results obtained from IBM Quantum hardware and compares them with the corresponding simulation data. The analysis focuses on the Entanglement Infidelity (EI) used primarily for MRB and the peak of doom, adapted from QA in this work. By examining the behavior of these quantities in both simulated and experimental settings, the robustness of the proposed framework and the influence of hardware noise on the observed correlation structures can be assessed following the validation.

\subsection{Validation of the MQA framework}

During the development of the MQA extension to the MRB framework, the initial objective was to verify that the introduction of the rotation layer does not significantly alter the underlying exponential decay behavior of MRB. However, early numerical simulations revealed that the decay-curve fitting became unstable at higher qubit numbers, with poor convergence and increased scatter in the extracted parameters.

To address this, two methodological refinements were introduced to make the analysis compatible with the MQA structure. Firstly, the parity-based measurement as discussed in Sec. \ref{sec:parameters} was adopted, replacing raw bitstring evaluation with XOR-based parity extraction to better capture the pairwise entanglement structure generated by MQA. Second, the initial and final single-qubit Clifford layers were omitted, as these layers can conjugate entangled pairs causing their parity to not always be well defined, preventing the use of the aforementioned parity method.

These modifications are motivated by the structure of the MQA entangling layer, which preserves qubit-pair parity and generates strong localized correlations. With these adjustments, the decay fitting becomes significantly more stable and physically consistent.

\begin{figure*}[p]
    \centering

    \includegraphics[width=0.9\textwidth]{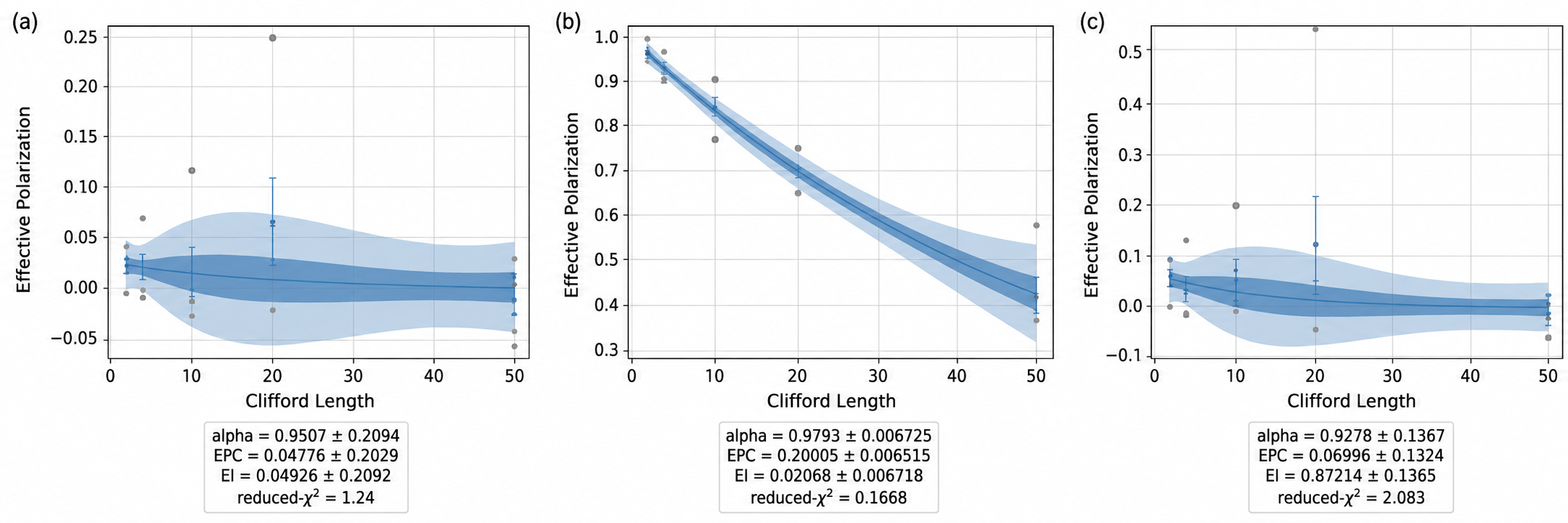}
    \caption{\textbf{Effective Polarization Fitting.} Comparison of EP decay for Mirror Quantum Awesomeness (MQA). All three curves were simulated on the same noisy backend using identical sequence lengths and random samples per length.
    (a) MQA analyzed with standard MRB and conventional bit-string analysis.
    (b) MQA without initial and final single-qubit Clifford layers, using parity-based measurement analysis.
    (c) MQA with initial and final Clifford layers.}
    \label{fig:mqa-refinements}

    \vspace{1em}

    \includegraphics[width=0.6\textwidth]{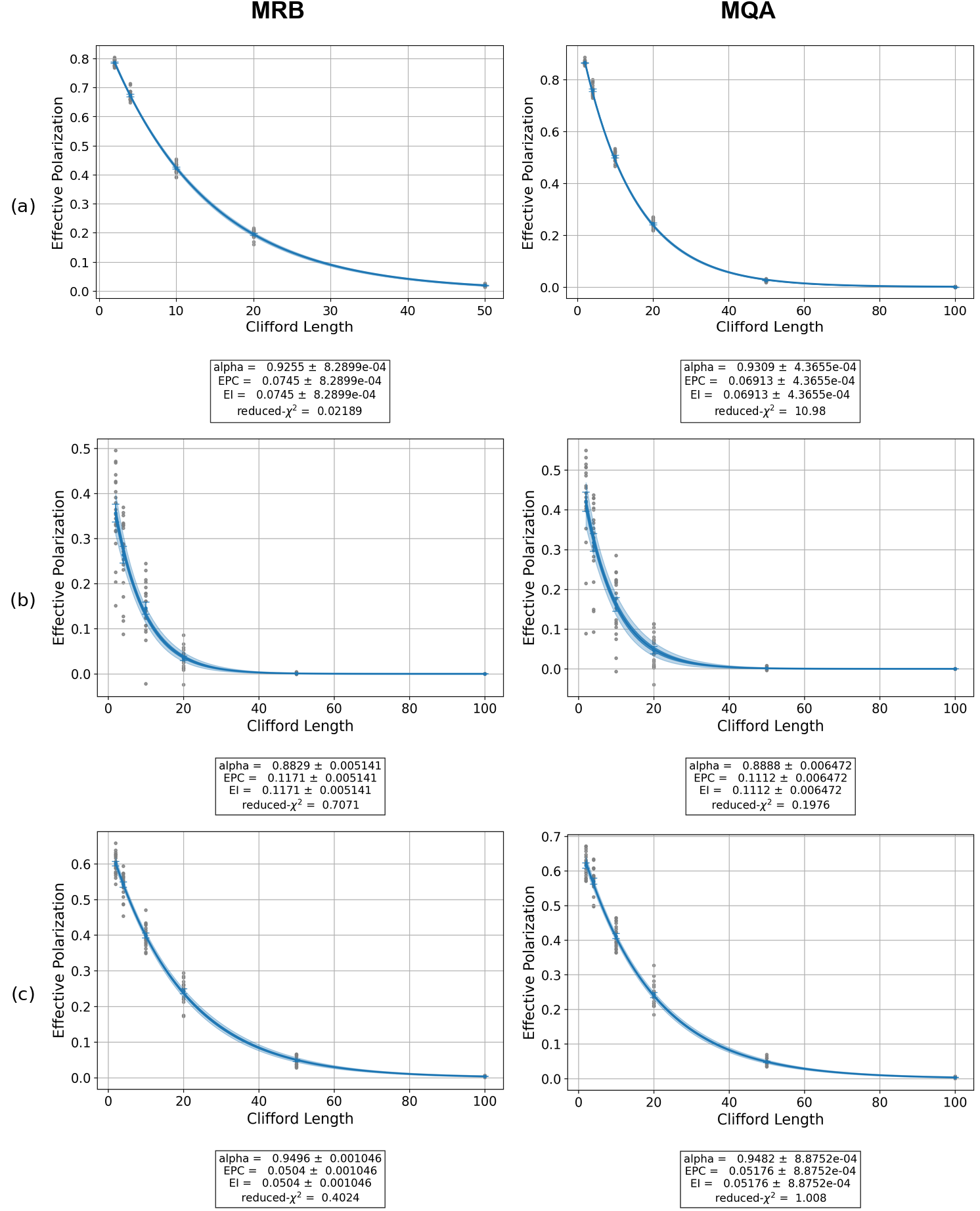}
    \caption{\textbf{MRB vs MQA Effective Polarization Comparison.} Comparison of EP graphs for the MRB and MQA protocols, with emphasis on the corresponding Entanglement Infidelity (EI) values. All results are generated under identical conditions, with a fixed number of \texttt{shots = 10000}, a constant qubit size of 20, and a fixed angle of $\theta = 0$. (a) Stabilizer backend. (b) \texttt{ibm\_fez}. (c) \texttt{ibm\_kingston}.}
    \label{fig:comparison-graphs}

\end{figure*}

As shown in Fig.~\ref{fig:mqa-refinements}, the MQA protocol without initial and final Clifford layers in the middle  produces the cleanest exponential decay, with the highest decay parameter $  \alpha  $ (closest to 1) and the lowest reduced $  \chi^2  $, indicating superior fit quality and stable EI extraction. In comparison, using standard MRB to analyze MQA results shown in Fig.~\ref{fig:mqa-refinements}'s left column shows noticeable scatter, while using MQA shown in Fig.~\ref{fig:mqa-refinements}'s right column shows additional variability due to fluctuations in parity. These corrections were then carried forward for all subsequent simulation and hardware evaluations to ensure a consistent and fair comparison framework.

\subsection{Entanglement Infidelity}

The EI analysis was performed across both numerical simulation and IBM quantum hardware to compare MRB and MQA under identical conditions.

In simulations, both protocols were evaluated using a stabilizer-based noisy 
simulator across system sizes scaling from $2$ to $156$ qubits. The error model 
incorporated depolarizing noise channels parameterized by single-qubit and 
two-qubit error rates of $p_1 = 0.001$ and $p_2 = 0.01$, respectively. Circuit 
executions were configured using $10,000$ shots with a constant sample size of 
$20$. To ensure strict architectural alignment with physical quantum hardware, 
both MQA and MRB were tested and compared using the heavy-hex coupling layout 
of the \texttt{ibm\_fez} device as a topological reference. Benchmarking and comparative 
validations were done across two operational regimes defined by $\theta = 0$ and $\theta = \pi/2$.

The comparative results demonstrate that the mean discrepancy in the EI values 
between the two methods remains bounded within $10\%$. Specifically, in the 
$\theta = \pi/2$ regime, the MQA method exhibits an average percentage 
discrepancy of approximately $-9.77\%$ relative to the baseline MRB benchmark. 
While individual variations were observed across different system sizes, the 
performance of MQA remains consistently within a narrow deviation range from 
the MRB standard.

\begin{table}[ht]
\centering
\caption{Comparison of EI values.}
\label{tab:ei_comparison}
\begin{tabular}{lccc}
\toprule
Environment & EI (MRB) & EI (MQA) & Difference (\%) \\
\midrule
Simulation   & 0.0745  & 0.06913 & -7.21 \\
\texttt{ibm\_fez}      & 0.1171  & 0.1112  & -5.04 \\
\texttt{ibm\_kingston} & 0.05040 & 0.05176 & +2.70 \\
\bottomrule
\end{tabular}
\end{table}

Similarly, on physical hardware, both MRB and MQA circuits were executed on the \texttt{ibm\_fez} and \texttt{ibm\_kingston} processors across a range of system scales, including 20, 50, 70, and the full device configuration. These physical evaluations were conducted under identical experimental parameters for both operational regimes  $\theta = 0$ and $\theta = \pi/2$ to enable a direct, comprehensive cross-platform comparison. The extracted EI values across these platforms demonstrate strong agreement, with structural deviations consistently remaining within the $10\%$ tolerance window. For clarity in visual presentation, the corresponding results and comparative curves for the 20-qubit configuration operating at  $\theta = 0$ are illustrated in Fig.~\ref{fig:comparison-graphs}.  The accompanying subplots further compare the Effective Polarization (EP) metric across these simulation and hardware platforms. To complement these graphical trends, the exact numerical values for all tested configurations are compiled in Table \ref{tab:ei_comparison}, providing a direct quantitative benchmark.

In both cases, the observed deviations between MQA and MRB remain small relative to the absolute EI values and are of the same order of magnitude. These results demonstrate that MQA provides EI estimates that are broadly consistent with those from standard MRB, with only modest protocol-dependent variations. 

\subsection{Peak of Doom}\label{sec:peak}

\begin{figure}[!htbp]
    \centering
    \includegraphics[width= \linewidth]{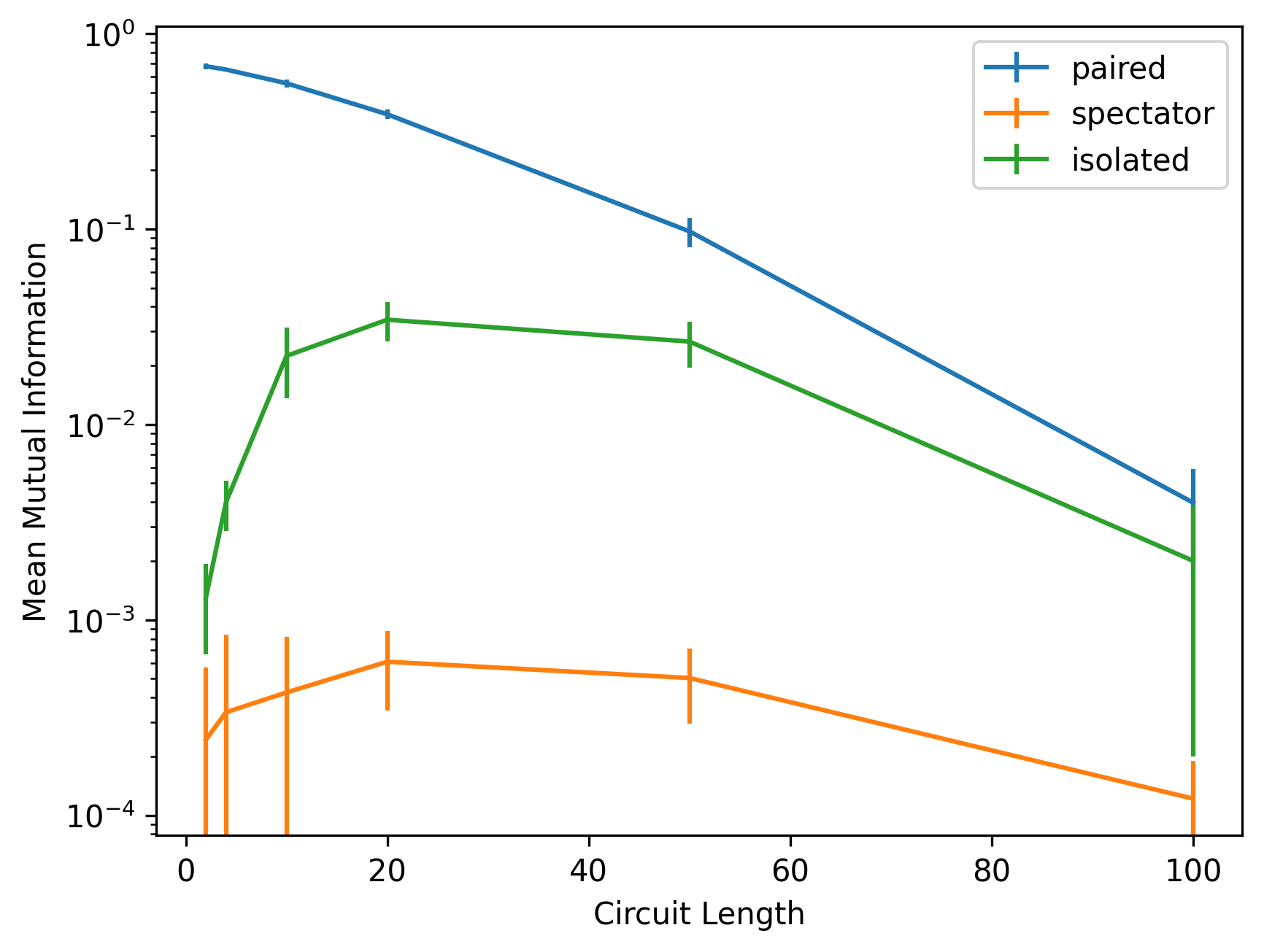}
    \caption{\textbf{MMI for \texttt{ibm\_fez} for $\theta = \pi/2$}. MQA experiments of 156 qubits on \texttt{ibm\_fez} with \texttt{samples = 20} and \texttt{shots = 10000}. The initial entangling angle is fixed to $\theta = \pi/2$.}
    \label{fig:mqa_real_fez_whole_pi_over_2}
\end{figure}

Fig.~\ref{fig:mqa_real_fez_whole_pi_over_2} shows how MQA performs on real hardware for  \texttt{ibm\_fez} with $\theta=\pi/2$. Similarly to what was observed for the customized noisy backend and the stabilizer backend, the spectator edges are unable to build correlations, and only the isolated edges show a peak. The real hardware therefore follows the expected behavior. The peak is largely a plateau between lengths 20 and 50, and the MI of the paired qubits decays to 0.1 at around length 50.

\begin{figure}[!htbp]
    \centering
    \includegraphics[width= \linewidth]{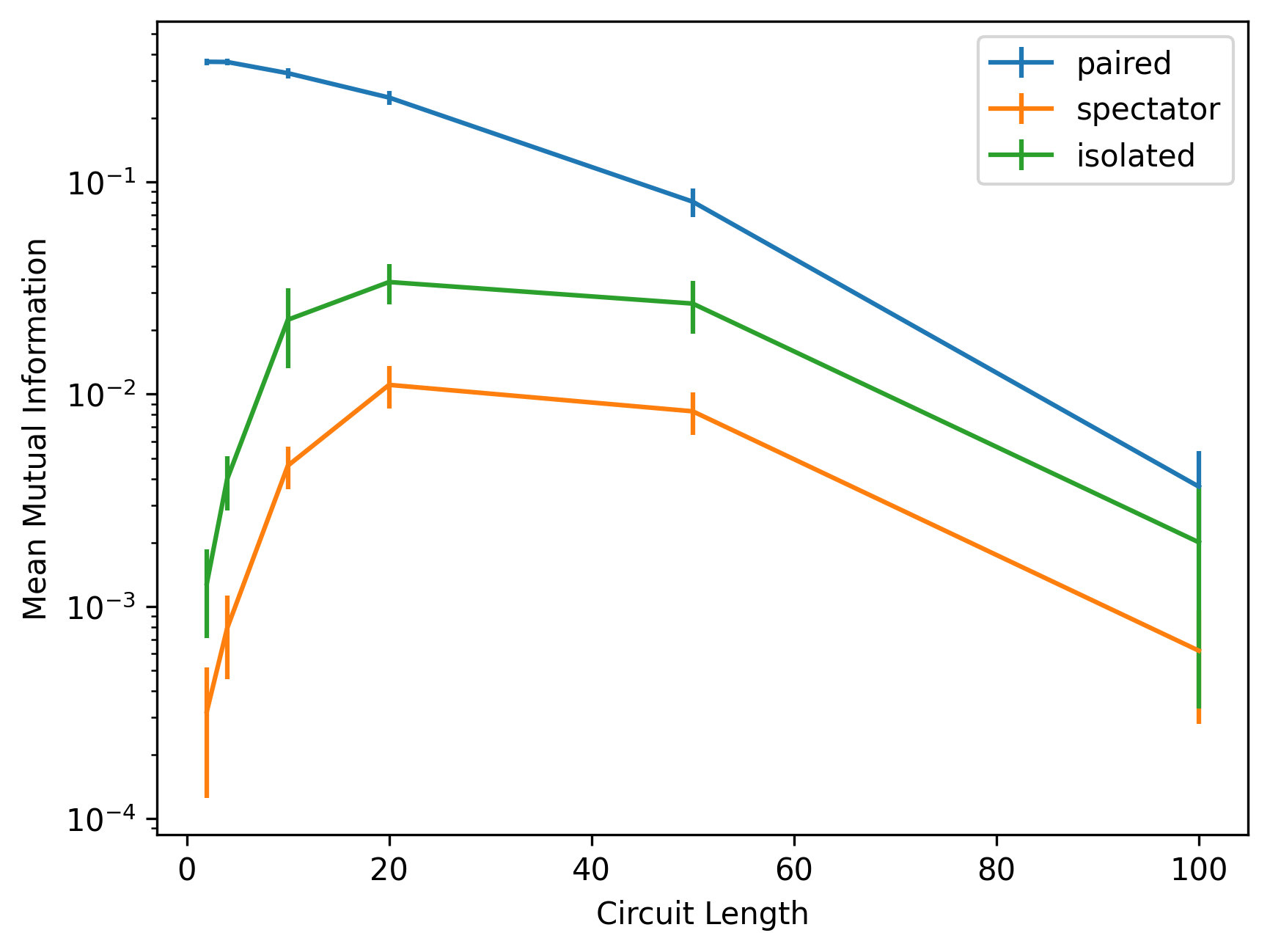}
    \caption{\textbf{MMI for \texttt{ibm\_fez} for $\theta = \pi/4$}. MQA experiments of 156 qubits on \texttt{ibm\_fez} with \texttt{samples = 20} and \texttt{shots = 10000}. The initial entangling angle is fixed to $\theta = \pi/4$.}
    \label{fig:mqa_real_fez_whole_pi_over_4}
\end{figure}

Fig.\ref{fig:mqa_real_fez_whole_pi_over_4} shows the same, but with the initial entangling angle set to $\pi/4$. This causes the spectator edges to experience correlation build-up as expected, once again matching the behavior observed with the custom noisy backend.

These results suggest that \texttt{ibm\_fez} operates at a noise level broadly compatible with the custom noisy simulator for $p_2=1\%$.

\subsection{TMQA Critical Point} \label{app:HW_topo}

\begin{figure}[!htbp]
     \centering
     \includegraphics[width= \linewidth]{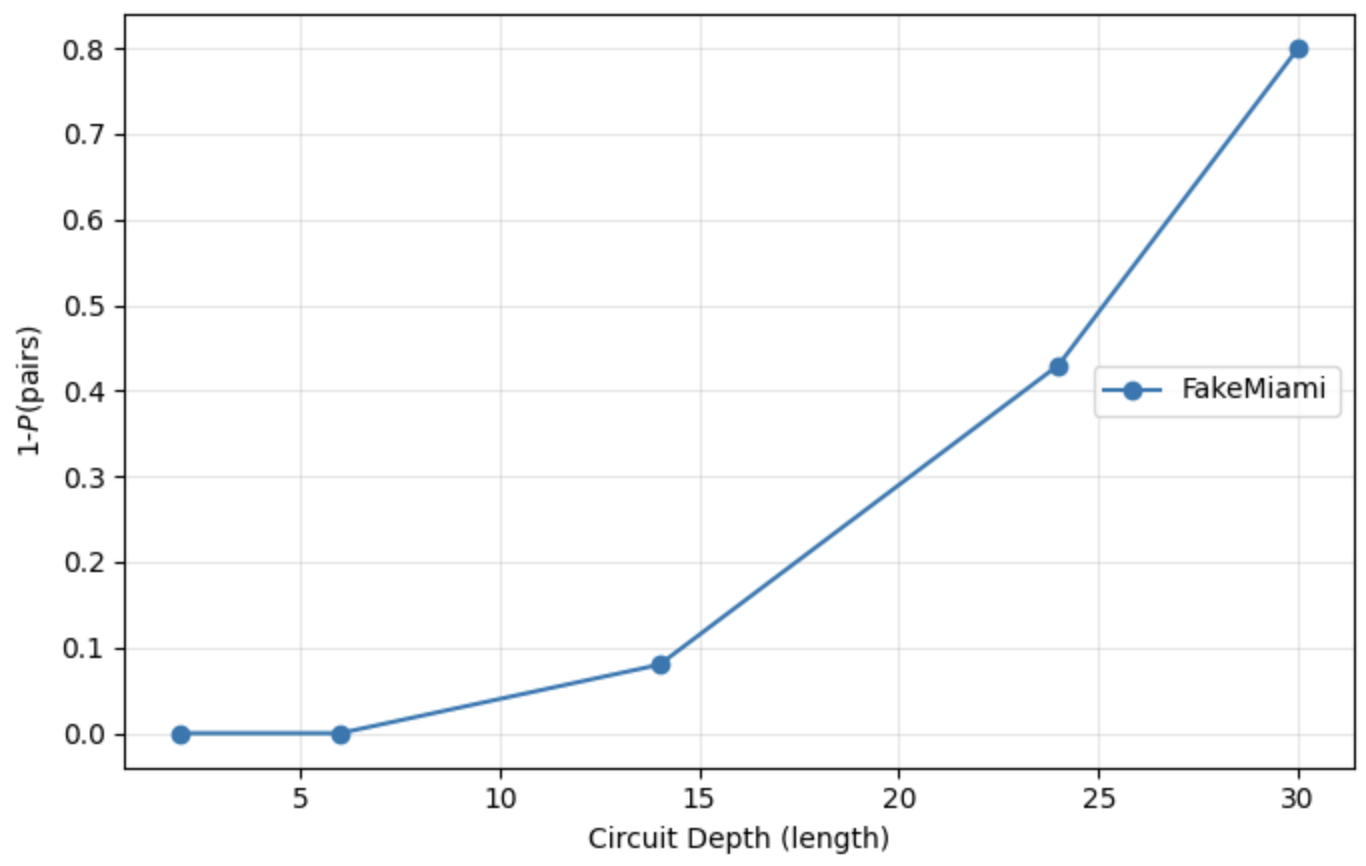}
     \caption{\textbf{TMQA for \texttt{FakeMiami}}. With \texttt{samples = 100} and \texttt{shots = 1000}. The initial entangling angle is fixed to $\theta = \pi/2$.}
     \label{fakemiami}
\end{figure}

\begin{figure}[!htbp]
     \centering
     \includegraphics[width= \linewidth]{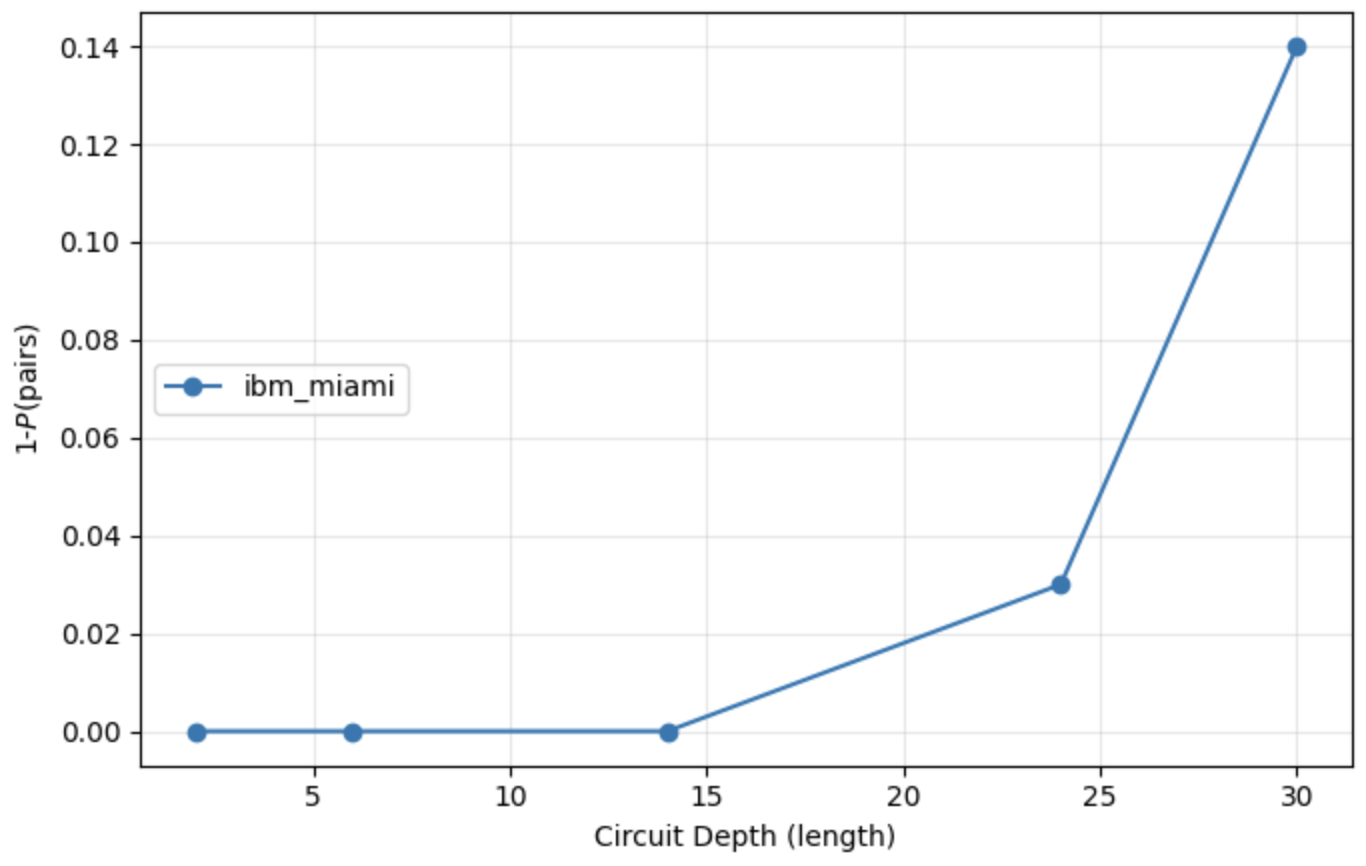}
     \caption{\textbf{TMQA for \texttt{ibm\_miami}}. With \texttt{samples = 100} and \texttt{shots = 1000}. The initial entangling angle is fixed to $\theta = \pi/2$.}
     \label{realmiami}
\end{figure}

For TMQA, a coupling graph is required that allows for all qubits to be paired (a perfect matching). This is not possible with the heavy-hex architecture of \texttt{ibm\_fez} and \texttt{ibm\_kingston}. However, it is compatible with the square lattice architecture of the 120 qubit \texttt{ibm\_miami}~\cite{ibmnighthawkmiami}. TMQA was therefore tested on this device, both for the real QPU and in simulation.

For simulation the \texttt{FakeMiami} backend was used. However, since several connections in the coupling map had a $100\%$ two-qubit gate error probability, these dead connections required modifications to the methods used to analyze the data. For $P_{\mathrm{pairs}}$ this was done by simply lowering the accuracy threshold: rather than being the probability that all pairs were correct, it was defined as the probability that at least $90\%$ of pairs were correct. This accounts for the number of observed dead connections with a small margin, allowing errors due to these dead connections to be easily ignored.

The plots of $P_{\mathrm{pairs}}$ in Fig.~\ref{fakemiami} and Fig.~\ref{realmiami} depict results from both the simulated and real backends, respectively. The results show remarkable agreement for the position of the critical depth. This shows that the noise model of the fake backend, which is informed by hardware calibration data, accurately captures qualitative features of the device's capabilities.

For $P_{\mathrm{topo}}$, no such simple modification is possible. The dead connections result in a coupling graph that does not admit a perfect matching. No clean results for this parameter could therefore be recovered.

\section{Conclusions}\label{sec:conclusions}

This work introduces Mirror Quantum Awesomeness and Topological Mirror Quantum Awesomeness. These are benchmarking protocols that extend Mirror Randomized Benchmarking by incorporating elements of Quantum Awesomeness, achieved by introducing a structured entangling rotation layer. Our results demonstrate that MQA preserves the reliable benchmarking capabilities of standard MRB while enabling enhanced characterization of the dynamics of entanglement and correlation structures.

Validation on a noisy qubit simulator confirmed that the inclusion of the MQA rotation layer produces decay curves consistent with conventional MRB, demonstrating that the base benchmarking behavior remains intact. Building upon this stable foundation, the introduction of a parity-based measurement analysis and the strategic omission of initial and final Clifford layers significantly enhanced fit quality.

On IBM Quantum hardware, MQA was run at scale on the \texttt{ibm\_fez} and \texttt{ibm\_kingston} processors across system sizes from 20 qubits to the full devices. The EI extracted from MQA agreed with that of MRB to within $10\%$ across all tested configurations (Table~\ref{tab:ei_comparison}), confirming that the MQA augmentation does not perturb the underlying MRB fidelity estimate. The peak of doom on the full 156-qubit \texttt{ibm\_fez} device was located at a depth of approximately $50$, consistent with the behavior of the depolarizing-noise reference model at $p_2 \approx 1\%$.

Taken together, the peak of doom from MQA and the threshold collapse from Topological MQA give two complementary methods for locating a device's critical circuit depth: the depth at which noise overwhelms the structured correlations the protocol attempts to preserve. These critical depths offer a practical guide to the circuit depth at which error-mitigation via pre- or post-processing alone might be expected to lose effectiveness, complementing the entanglement infidelity as a per-device performance metric.

\section*{Acknowledgments}

This work was supported as a part of NCCR SPIN, a National Centre of Competence in Research, funded by the Swiss National Science Foundation (grant number 225153).

The authors acknowledge the use of IBM Quantum Credits via the IBM Quantum Startups Program for this work. The views expressed are those of the authors and do not reflect the official policy or position of IBM or the IBM Quantum Platform team.

\vspace{1em}

\textbf{Data Availability}. The data and code used for all experiments presented in this work can be found in \cite{mqa-data}.

\clearpage

\bibliography{references}

\end{document}